\documentclass[journal,twoside,web]{ieeecolor}

\usepackage{generic}
\usepackage{cite}
\usepackage{amsmath,amssymb,amsfonts}
\usepackage{algorithmic}
\usepackage{graphicx}
\usepackage{textcomp}
\def\BibTeX{{\rm B\kern-.05em{\sc i\kern-.025em b}\kern-.08em
    T\kern-.1667em\lower.7ex\hbox{E}\kern-.125emX}}
\begin{document}
\title{Analysis of Intra-Operative Physiological Responses Through Complex Higher-Order SVD for Long-Term Post-Operative Pain Prediction}
\author{Raheleh Baharloo, ~Jos\'{e} C. Pr\'{i}ncipe,~\IEEEmembership{Fellow,~IEEE}, Parisa Rashidi, and Patrick J. Tighe 
\thanks{``This work was supported in part by the National Institute of General Medical Sciences under Grant 5R01GM114290 and the National Institute of Aging under Grant 5R01AG055337.'' }
\thanks{R.~Baharloo and J.~C.~Pr\'{i}ncipe are with the Electrical and Computer Engineering Department, University of Florida, Gainesville, FL 32611, USA (e-mail: baharloo@ufl.edu; principe@cnel.ufl.edu). }
\thanks{P.~Rashidi is with the Biomedical Engineering Department, University of Florida, Gainesville, FL 32611 ,USA (e-mail: parisa.rashidi@bme.ufl.edu).}
\thanks{P. J. Tighe is with 
the Anesthesiology Department, University of Florida, Gainesville, FL 32611 ,USA (e-mail: ptighe@ufl.edu).}}
\maketitle

\begin{abstract}
Long-term pain conditions after surgery and patients' responses to pain relief medications are not yet fully understood. While recent studies developed an index for nociception level of patients under general anesthesia, based upon multiple physiological parameters, it remains unclear whether and how the dynamics of these parameters indicate long-term post-operative pain. To extract unbiased and interpretable descriptions of how physiological parameters dynamics change over time and across patients in response to surgical procedures and intra-operative medications, we employed a multivariate-temporal analysis. We demonstrate the main features of intra-operative physiological responses can be used to predict long-term post-operative pain. 
We propose to use a complex higher-order SVD (complex-HOSVD) method to accurately decompose the patients' physiological responses into multivariate structures evolving in time. We used intra-operative vital signs of 175 patients from a mixed surgical cohort to extract three interconnected, low-dimensional complex-valued descriptions of patients' physiological responses: multivariate factors, reflecting sub-physiological parameters; temporal factors reflecting common intra-surgery temporal dynamics; and patients factors, describing patient to patient changes in physiological responses. Adoption of complex-HOSVD allowed us to clarify the dynamic correlation structure included in intra-operative physiological responses. Instantaneous phases of the complex-valued physiological responses within the subspace of principal descriptors enabled us to discriminate between mild versus severe levels of pain at post-operative Day-30 and Day-90.  By abstracting patients into different surgical groups, we identified significant surgery-related principal descriptors: each of them potentially encodes different surgical stimulation. The dynamics of patients' physiological responses to these surgical events are linked to long-term post-operative pain development.
 
\end{abstract}


\section{Introduction}
\label{sec:introduction}
\IEEEPARstart{P}{ersistent}  pain after acute post-operative pain (POP) is experienced by 10-50\% of individuals after common surgical procedures like the cardiac, thoracic, spine or orthopedic surgeries ~\cite{kehlet2006persistent}. Although even mild levels of persistent post-operative pain (POP) are associated with decreased physical and social activities ~\cite{haythornthwaite1998pain}, 2-10 \% of patients experiencing this type of pain may develop severe levels of pain, hence delaying recovery and their return to the normal daily function \cite{perkins2000chronic,macrae2001chronic}. Furthermore, persistent POP leads to increased direct medical costs through additional resource use. Persistent POP appears to be a critical, mainly unrecognized clinical problem ~\cite{kehlet2006persistent}. Consequently, recognition of patients at risk of developing this type of pain has remained inadequate ~\cite{benhamou2008postoperative}. \\
\indent 
\begin{figure*}[h!]
\centering
\includegraphics[width=0.85\textwidth]{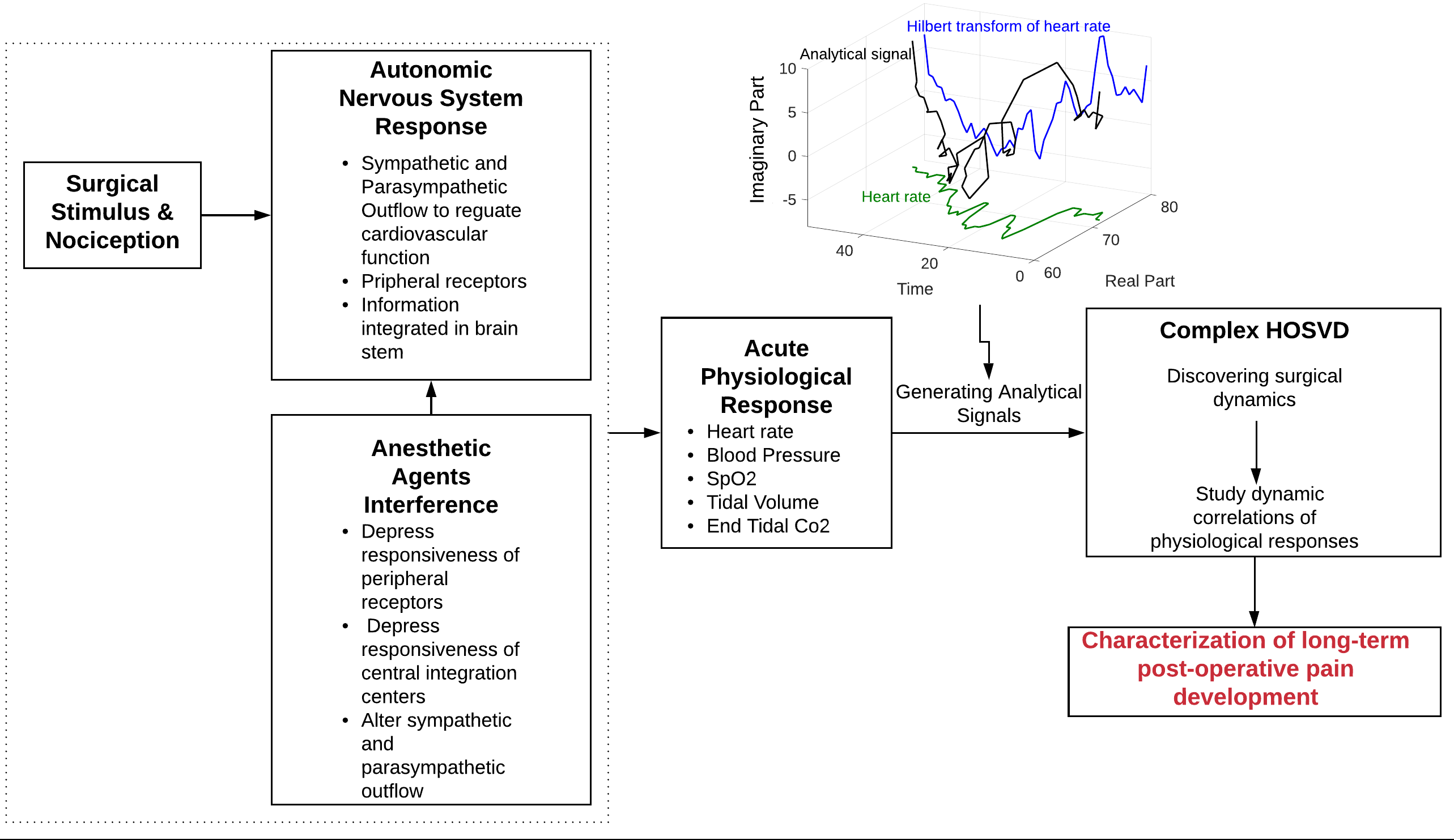}
\caption{Flow diagram of the work proposed in this study. Multivariate intra-operative vital signs as indicators of the dynamic interplay among surgical stimulus, autonomic nervous system, and anesthetic agents are analyzed through tensor decomposition to characterize long-term post-operative pain.} 
\label{Fig:1}
\end{figure*}
POP is assumed to stem from various interacting factors including but not limited to biological, psychological, and social factors ~\cite{turk1999role}. In different studies ~\cite{hinrichs2009psychosocial,sullivan2011role}, psychological factors (depression, psychological vulnerability, stress, and catastrophizing) are suggested as risk factors for the development of persistent POP. Level of education and female gender were seen by some as unlikely to be coupled with persistent POP ~\cite{hinrichs2009psychosocial}. However, Holtzman et al. ~\cite{holtzman2002gender} identified female sex as a risk factor for developing persistent POP. The relationship between anxiety and the development of persistent POP is not yet clear. While various studies discovered a significant link between pre-operative anxiety and higher levels of persistent POP ~\cite{gerbershagen2009risk,pinto2012risk}, several studies did not discover this association. Hinrichs-Rocker et al. ~\cite{hinrichs2009psychosocial} in a systematic review evaluating the association between anxiety and persistent POP in patients undergoing different types of surgery, found no clear linkage between the two. In a meta-analysis evaluating 29 different research studies, Theunissen et al.  ~\cite{theunissen2012preoperative} said preoperative anxiety was remarkably associated with persistent POP in only 55 \% of the studies. 

A frequently replicated finding suggests that the severity of acute POP ~\cite{gehling1999persistent,bisgaard2005acute,kehlet2006persistent}, especially movement-evoked pain ~\cite{katz1996acute,tasmuth1997treatment,callesen1999prospective} is the most striking risk factor significantly associated with persistent POP. Basbaum ~\cite{basbaum1999spinal} said neuroplastic changes in the central nervous system resulting from high intensities of acute POP were a reason for developing persistent pain. In all these studies on the impact of acute POP on the development of persistent POP, one single measurement of acute pain (mean daily value, or the worst one) was examined, and the time resolution of acute pain was discarded. 

In recent years, the acute POP dynamic (POP trajectory) as quantification of all apparent and latent factors controlling post-operative pain duration and resolution was examined through different methods to identify abnormal POP resolution ~\cite{chapman2011improving,althaus2014distinguishing}. Chapman approximated daily pain trajectories using a linear mixed model to increase the information extracted from post-operative pain recordings. Through their method, three pain trajectory patterns unfolded, yielding new information about the dynamics of post-operative pain resolution in a limited time window after surgery. Later, Althaus et al. ~\cite{althaus2014distinguishing} used a latent growth curve on the average pain intensities over the first 5 days after surgery to analyze the mediating effects of post-operative pain trajectories within the association between some relevant pre-operative psychosocial features and chronic post-surgical pain. Notably these extensions to pain trajectories generally focused on daily abstractions of pain intensity ratings and discarded potentially meaningful data pertaining to intra-day variations. 
Furthermore, they used constrained models to approximate the complex dynamic of post-operative pain resolution. Baharloo et al. ~\cite{baharloo2021slow} extended this line of research by considering POP intensity observations including intra-day variations, as a time series, and used wavelets to approximate the POP temporal dynamics associated with persistent POP. 

Although these studies are encouraging, their strategies are inherently limited by a lack of analysis of nociception, that is, the sensory nervous system response to harmful or potentially harmful stimuli. Every individual may respond differently to a painful stimulus, and the characteristics of this response may indicate further development of persistent pain. Hence, we argue that to find a solution for this complex problem, one needs to carefully analyze the inherent response to a painful stimulus, to characterize the intricate nature of persistent pain.  In this study, we considered the systematic physiological response to surgical injury to study the reaction of every individual in response to a noxious stimulus.

During surgery, as the autonomic nervous system continuously responds to various surgical stimuli, vital signs such as heart rate, blood pressure, and respiration can be used as indicators of these responses. During general anesthesia, when a sufficient dose of anesthetic agent is applied to prevent the response to skin incision, physiological responses induced by surgical stress are not necessarily attenuated~\cite{kazama1998pharmacodynamic}. The sympathetic nervous system inherently changes physiological parameters such as local blood flow, blood pressure, and heart rate in response to noxious stimulation. Anesthetic agents do interfere with this system at different levels. Among physiological parameters, heart rate may also include changes in parasympathetic discharge~\cite{guyton1951pressoreceptor}. Hence, monitoring and analyzing the time series of patients’ physiological responses in relation to a variety of surgical stimuli and nociception imbalance under general anesthesia helps us to indirectly characterize the behavior of the autonomic nervous system to nociceptive stimuli and will potentially provide a clue for the development of persistent POP.

Hemodynamic regulation is the result of dynamic interactions between coupled biological systems of different scales and temporal frequencies. The cross-spectral analysis is a proper method to inspect such phenomena. However, when dominant frequencies and scales are unknown or happen over a wide range, employing cross-spectral analysis in an exploratory manner is difficult. Furthermore, when dealing with non-stationary time series characterized by short-time and irregularly occurring events, as is the case in intra-operative vital signs, cross-spectral analysis is less descriptive.

Matrix (second-order tensor) singular value decomposition  (SVD) is a mathematical tool for rank-revealing, dimensionality reduction, and subspace detection ~\cite{jiang2000solving}. The method has been extensively used to explore multivariate and temporal relationships available in multivariate time series data sets. The main interest of the approach is its capability to capture pairwise interactions and to find the fewest number of orthogonal components to explain the complex variability of the original time series. For applications requiring higher-order tensors, a proper generalization of the matrix SVD called higher-order SVD (HOSVD) gaining from the power of multilinear algebra is introduced by De Lathauwer et al. ~\cite{de2000multilinear}. The HOSVD approach captures multiple interactions and couplings within multiway data and discovers multidimensional patterns/dynamics. When employed on intra-operative vital signs, the method is potentially able to learn the low-dimensional representation of the surgical stimulus space (surgical dynamics) shared by different physiological responses. 

However, both matrix SVD and HOSVD approaches if employed in conventional form, detect standing dynamics and oscillations, not propagating one. The methods can be extended in different ways to detect propagating dynamics. If a time series is the only obtainable measurement, stacking multiple time-lagged duplicates of the series into an augmented data matrix/tensor help to capture the phase information of the underlying dynamics. However, performing matrix SVD/HOSVD on the augmented matrices/tensors only resolve propagating dynamics over a restricted time span. 

Singular value decomposition can be performed in the frequency domain to study propagating dynamics. As analogous to principal component analysis in the frequency domain (FDPC) ~\cite{wallace1972empirical}, the method involves calculating complex eigenvectors from cross-spectral matrices/tensors. Due to the stochastic nature of physiological responses, employing the methods in the frequency domain may be complicated. For example, the phase information of the series may vary drastically in successive windows of time, or if the power of an eigenvector is spread over multiple frequencies, then many phase relationships (one for each spectral estimate) must be examined to  accurately study the propagating dynamics.   

Complex principal component analysis in the time domain (CPCA)~\cite{horel1984complex} is an alternative approach to FDPC. In this method, the real-valued time series are complexified using their Hilbert transform. Subsequently, complex eigenvectors are calculated from cross-covariance/cross-correlation matrices obtained from the complexified time series. CPCA analysis is primarily FDPC analysis averaged over all frequency bands ~\cite{brillinger2001time}. Both methods are identical if the underlying dynamic of a time series is governed by the variability of one frequency only ~\cite{brillinger2001time}. Similar to CPCA,  matrix SVD/HOSVD can be employed on the complexified version of real-valued physiological responses (hereafter referred to as complex-matrix SVD/complex-HOSVD) to allow for   effective capturing of propagating dynamics while the power of the responses is spread over different frequency bands. 

Our study employs complex-HOSVD to explore dynamic correlations with lead/lag relations in intra-operative vital signs. The complex-valued vital signs are generated using the original ones and their Hilbert transforms. The key idea is to organize complex-valued vital signs into a third-order tensor with three axes corresponding to individual vital signs (physiological parameters), time during surgery, and patients. We then fit the HOSVD to identify a set of low-dimensional complex-valued factors (features) capturing variability along each of these three axes. 

The complex-HOSVD identifies separate low-dimensional complex-valued factors, each of which corresponds to sub-physiological parameters with common within-surgery dynamics and variable across-patient dynamics. We then investigate how surgical mechanisms in different surgeries emerge over the process of physiological responses. The investigation elucidates that each surgical service has its particular dynamics captured in some significant factors with different characteristics. We discuss that the complex-HOSVD can extract descriptors of physiological responses in which individual factors potentially correspond with interpretable activities such as tidal volume determination and autonomic regulation during surgery.     

Finally, We employ the complex-valued factors as new bases to describe physiological responses.  After projection onto the subspace, the complex correlations between each intra-operative time series and the complex-valued factors are manifested in magnitudes and phases of the correlations. We use the phases of the correlations to predict mild versus severe levels of pain on post-operative Day-30 and Day-90. We demonstrate dissimilarities between these two pain categories are relatively expressed in the phase information of the physiological responses with respect to surgical dynamics. Figure \ref{Fig:1} shows the relation of the proposed tasks with the underlying biological sub-systems.
\\
\indent 
\begin{figure*}[h!]
\includegraphics[width=0.9\textwidth]{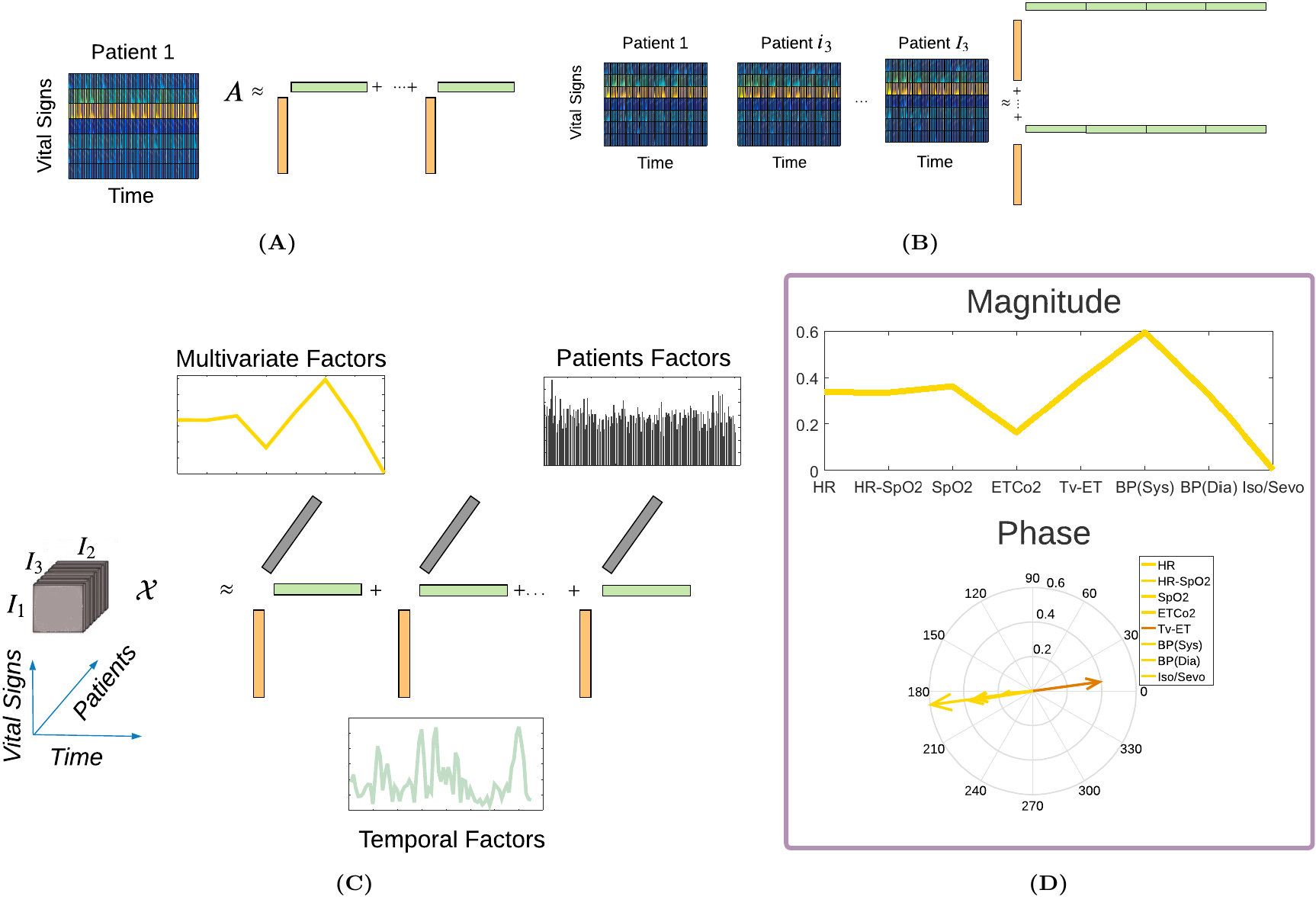}
\centering
\caption{Tensor representation of intra-operative vital signs} 
\label{Fig:2}
\end{figure*}
\section{Discovering surgical multivariate-temporal dynamics through complex-HOSVD}
\subsection{Intra-Operative Vital Sign Recording} 
During surgery, a patient experiences various disturbances with respect to the normal activities of different body systems. Therefore, monitoring the patient’s physiological status is essential. Physiologic monitoring systems can continuously measure and monitor various vital signs through electrodes and sensors connected to the patient. Routinely measured vital signs may contain the electrical activity of the heart (through an ECG), heart rate, respiratory rate, blood pressure, cardiac output, body temperature, SpO2, Et-CO2, and tidal volume (Tv-ET). 

\textbf{Blood pressure} is the pressure generated by circulating blood on the walls of blood vessels and usually points to the pressure in the large arteries. Blood pressure is commonly expressed as systolic and diastolic pressure. Systolic pressure refers to the amount of pressure in the arteries when the heart contracts to pump blood into circulation. Diastolic pressure refers to the pressure when the heart relaxes after contraction. During surgery, blood pressure can be measured by both invasive and non-invasive methods. Invasive monitoring of blood pressure involves direct estimation of arterial pressure by inserting a cannula needle into an appropriate artery. This provides continuous beat-by-beat monitoring of the patient’s blood pressure. Non-invasive monitoring uses an oscillometric technique with an automated cuff.

\textbf{SpO2} stands for peripheral capillary oxygen saturation and is measured noninvasively using a pulse oximeter, to provide an approximation of the arterial hemoglobin oxygen saturation. A sensor is clipped over the finger, and the pulse oximeter continuously emits and absorbs a light wave going through capillaries. Since the oxygen saturation level causes changes in the blood’s color, variations of the light wave provide an estimate for the value of the SpO2. The pulse oximeter also provides heart rate in beats per minute, as an average rate over 5 to 20 seconds.

\textbf{EtCO2} stands for end-tidal carbon dioxide and represents the amount of carbon dioxide in exhaled air. Tidal volume represents the volume of air displaced between inhalation and exhalation. These two parameters can be used to assess ventilation. 

\textbf{Isoflurane and Sevoflurane} (both included in the same class of chemicals) were used as two inhaled anesthetic agents in this study, and they both are known to have a depressive effect on the autonomic nervous system. The end-tidal concentration of Isoflurane/Sevoflurane is a clinical index to predict emergence from anesthesia. It is related to the sympathetic and parasympathetic drives in patients during surgery.

In this study, we used seven vital parameters including heart rate, heart rate-SpO2, SpO2, systolic blood pressure, diastolic blood pressure, ETCo2, Tv-Et, and end-tidal concentration of Isoflurane/Sevoflurane as superficial and imperfect indicators of autonomic nervous system activity or state. These parameters were subjected to analysis through tensor decomposition to characterize long-term post-operative pain.

\textbf{Long-term post-operative pain }is a self-reported mean value of pain on post-operative Day 30 using a numeric rating scale (0 no pain, 10 the worst pain). Although this method is not an ideal assessment of pain, and its potential subjective bias makes it less reliable, different studies have reported a significant correlation between this method and the pain measured by different candidate technologies such as physiological parameters or cerebral hemodynamic changes for pain assessments ~\cite{ben2013monitoring}.

\subsection{Application of SVD to Large-Scale Intra-Operative Data Analysis}
Before describing complex HOSVD in our analysis, we first discuss the potential application of SVD to larg-scale intra-operative data analysis. Consider a recording of $I_1$ intra-operative vital signs over $I_3$ different patients. We assume vital-signs are recorded at $I_2$ time points for each patient, but recordings of variable duration can be cut to a common window of time to fit in with this constraint. The collection of these series is naturally represented as an $I_1 \times I_2 \times I_3$ array of vital signs, a third-order tensor such as $ \mathcal{A} \in \mathbb{R}^{I_1 \times I_2 \times I_3}$ . Each member of this tensor, $a_{i_1 i_2 i_3}$ , denotes the recorded value of vital-sign $i_1$ at time point $i_2$ for patient $i_3$. 

Given that patients are under various surgical stimuli, such large multiway arrays (tensors) are challenging to analyze and interpret, particularly when the recordings are done over a wide range of surgical services. Even physiological responses to an identical surgical stimulus exhibit significant patient to patient variability.

Under the assumption of having the intra-operatove vital signs recorded  for just one patient, we obtain a matrix, $ \mathbf{A}_{I_1 I_2}$, which holds the values for each vital sign $i_1$ and time point $i_2$. Even such a matrix is hard to interpret when different vital signs with distinct temporal dynamics are involved in the experiments.

SVD summarizes this matrix by carrying out a decomposition into $R$ number of ranked-one matrices (components) such as in equation \ref{eqn:1} to approximate the original data matrix.
\begin{equation}
\mathbf{A}=\sum\limits_{r=1}^{R} \sigma_r \mathbf{A}_r;  \; \; \; \; \; \; \;  \; \; \; \; \; \; \; \; \;\mathbf{A}_r=U_r \circ V_r.  
 \label{eqn:1}
\end{equation}
where $\circ$ denotes the outer product of the vectors. This decomposition provides a low-dimensional subspace (a new coordinate system) with $R$ dimensions to describe the original high-dimensional data with $I_1$ or $I_2$ dimensions. Whenever decomposition applies, we will use the terms: dimensions, components, and ranked-one matrix/tensor interchangeably, but they convey the same meaning. Each ranked-one matrix, indexed by $r$, holds  a coefficient across vital signs,$u_{r{i_1}}$, and  a coefficient across points in time $v_{r{i_2}}$. These coefficients can be accumulated into vectors $U_r$ with length $I_1$ and $V_r$ with length $I_2$. These vectors represent the multivariate-temporal dynamics discovered within the original data matrix. In this paper, we call vectors $U_r$ as multivariate modes (factors), and vectors $V_r$ as temporal modes (factors)(yellow and green vectors in figure \ref{Fig:2} A, respectively). Each coefficient (element) of the multivariate (or temporal) mode vectors contains two important pieces of information. The absolute value of the coefficient provides a measure of the particular vital sign's (or time point's) contribution for that mode. If the coefficient is complex valued (as is the case with this paper), the angle defined by the real and imaginary parts provides an explanation of the phase of that coefficient (element) in relation to the others vibrating at the frequency associated with that particular mode ~\cite{kutz2016dynamic}.  
In order to account for variability of vital signs among patients and also simplify the data tensor $\mathcal{A}$, one approach is to concatenate multiple data matrices like $ \mathbf{A}_{I_1 I_2}$ (one for each patient), by that means converting the data tensor into an $I_1 \times I_2 I_3$ matrix, and then applying SVD to this matrix (Figure \ref{Fig:2} B). In this way, the $R$ temporal modes are of length $I_2 I_3$ and do not capture common temporal dynamics across patients.

In this paper, we perform decomposition directly on the original data tensor $\mathcal{A}$ (Figure \ref{Fig:2} C), rather than transforming it to a matrix . The HOSVD method then provides the following decomposition, 

\begin{equation}
\mathcal{A}=\sum\limits_{i_1} \sum\limits_{i_2} \sum\limits_{i_3} s_{i_1 i_2 i_3} U_{i_1}^{(1)} \circ U_{i_2}^{(2)} \circ U_{i_3}^{(3)} .  
 \label{eqn:2}
\end{equation}

In analogy to SVD, we can think of $U^{(1)}$ as a prototypical pattern across intra-operative vital signs, and $U^{(2)}$ as a temporal dynamic across time. These multivariate modes and temporal modes represent dynamics that are common among all patients. The third set of modes, $U^{(3)}$,"patients factors" (Figure \ref{Fig:2} C), represents patients-specific variations for the multivariate-temporal dynamics identified by the method.     

Furthermore, to capture propagating dynamics, the real-valued vital signs are augmented with their Hilbert transforms to form a complex-valued third-order tensor such as $ \mathcal{X} \in \mathbb{C}^{I_1 \times I_2 \times I_3}$. The HOSVD decomposition in equation \ref{eqn:2} holds for the complex-valued tensor $\mathcal{X}$ as well \cite{de2000multilinear}. The complex HOSVD identifies dynamic factors that carry additional information related to phase.  Figure \ref{Fig:2} D illustrates a single multivariate factor plotted with respect to magnitude and phase, where each element of the the multivariate factor represents a particular vital sign recorded during surgery.  The phase is plotted between 0 and 2$\pi$, representing the relative phases of the elements. The phase plotted on a circular grid exhibits an interesting feature. All elements of the multivariate factor shows the same phase except the element associated with contribution of tidal volume. Tidal volume is selected by the anesthesiologist during surgery, but still influences heart rate and blood pressure in patients.

\begin{figure}[!htbp]
\includegraphics[width=0.50\textwidth]{{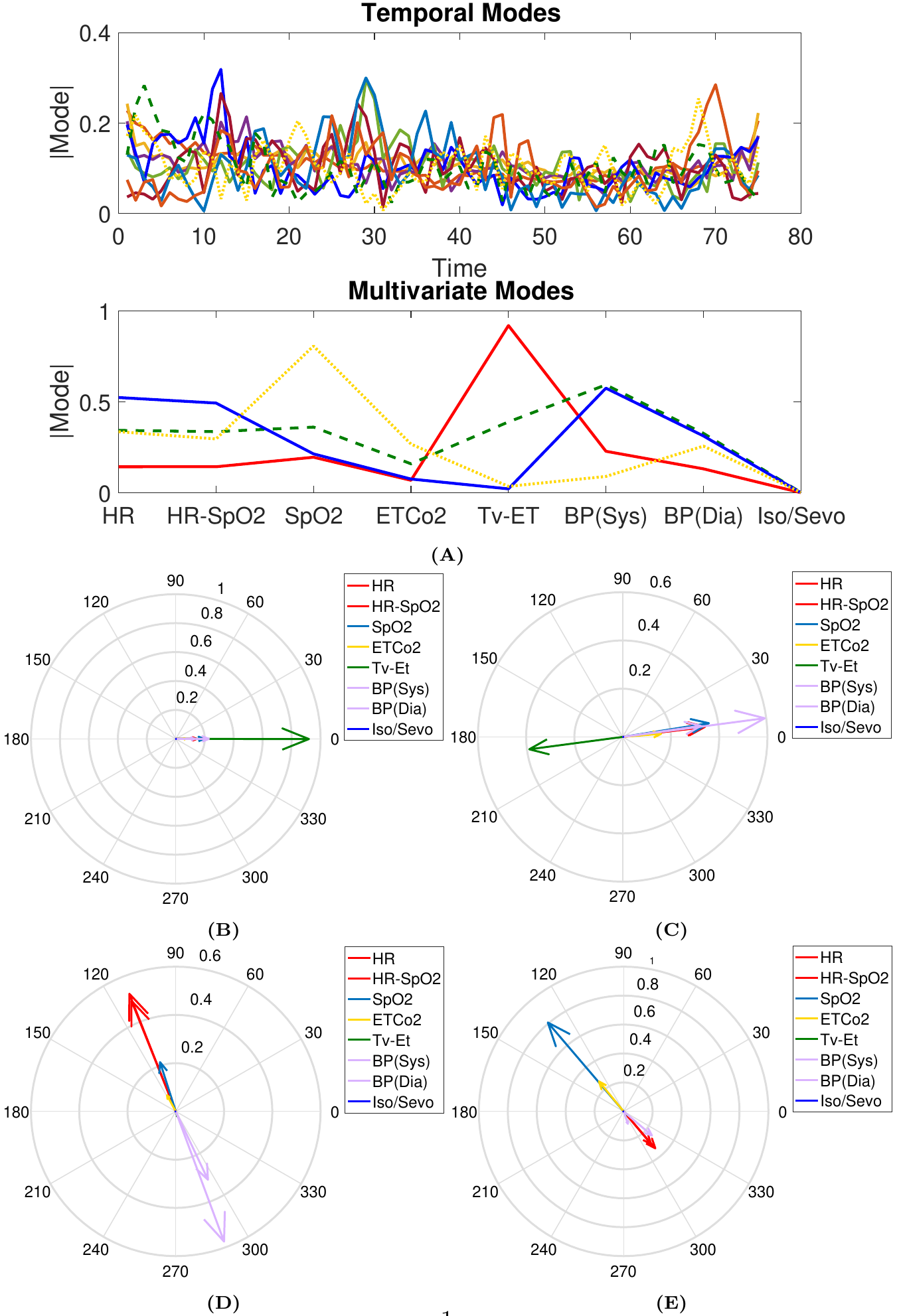}}

\caption{Illustration of the multivariate and temporal factors extracted through Complex HOSVD. A) The outer product of the temporal and multivariate factors generates the contributing components in the decomposition of $\mathcal{X}$. For intra-operative vital signs, each of the elements of a multivariate mode represents a particular vital sign. The magnitude and phase of the element explain how the vital signs are related to each other within that factor. The phase of each element describes the relative phase of the vital sign's vibration relative to the other vital signs for that multivariate factor.  This depiction allows for interpretation of the Complex-HOSVD output for intra-operative vital signs: each multivariate factor identifies the vital signs involved in that pattern of physiological response in addition to the relative phase of that vital sign's activation time. B, C, D, and E) represent the phase portrait associated with the multivariate factors shown in red, blue, green, and yellow, respectively.} 
\label{Fig:3}
\end{figure}

\subsection{Experimental Setup, Discovered Surgical Dynamics}
This prospective cohort study was approved by the University of Florida IRB-01 (Protocol no. IRB201500153) as the National Institutes of Health-funded TEMPOS protocol. A complex-HOSVD based metric projection is designed and tested in this study to characterize the individuals’ physiological dynamics from intra-operative vital signs collected during surgery with a rate of one sample per minute for at least 75 minutes.  The intra-operative vital signs are augmented by their Hilbert transform \cite{horel1984complex} to create the complex vital signs.  Here, a two-dimensional tensor (Matrix) represents the time-varying dynamics of different intra-operative vital signs for each patient.  The total 175 second-order tensors constructed from the intra-operative vital signs of 175 patients undergoing a relatively wide range of surgical services including orthopedic, urology, colorectal, transplant, pancreas and biliary, and thoracic surgeries, were stacked on top of each other to generate a three-dimensional tensor to discover mixed surgical dynamics. The complex principal multivariate and temporal factors extracted through complex-HOSVD were compared with the multivariate temporal dynamics extracted after grouping patients based on their surgical service. The complex-HOSVD decomposition resulted in the approximation of multivariate, and temporal factors with sizes $8 \times 4$, $75 \times 32$, respectively. Thus, a total of 128 ($F=4 \times 32$) features were extracted among 600 features of the tensors. \\
\indent 
\begin{figure*}[!htbp]
\centering
\includegraphics[width=0.9\textwidth]{{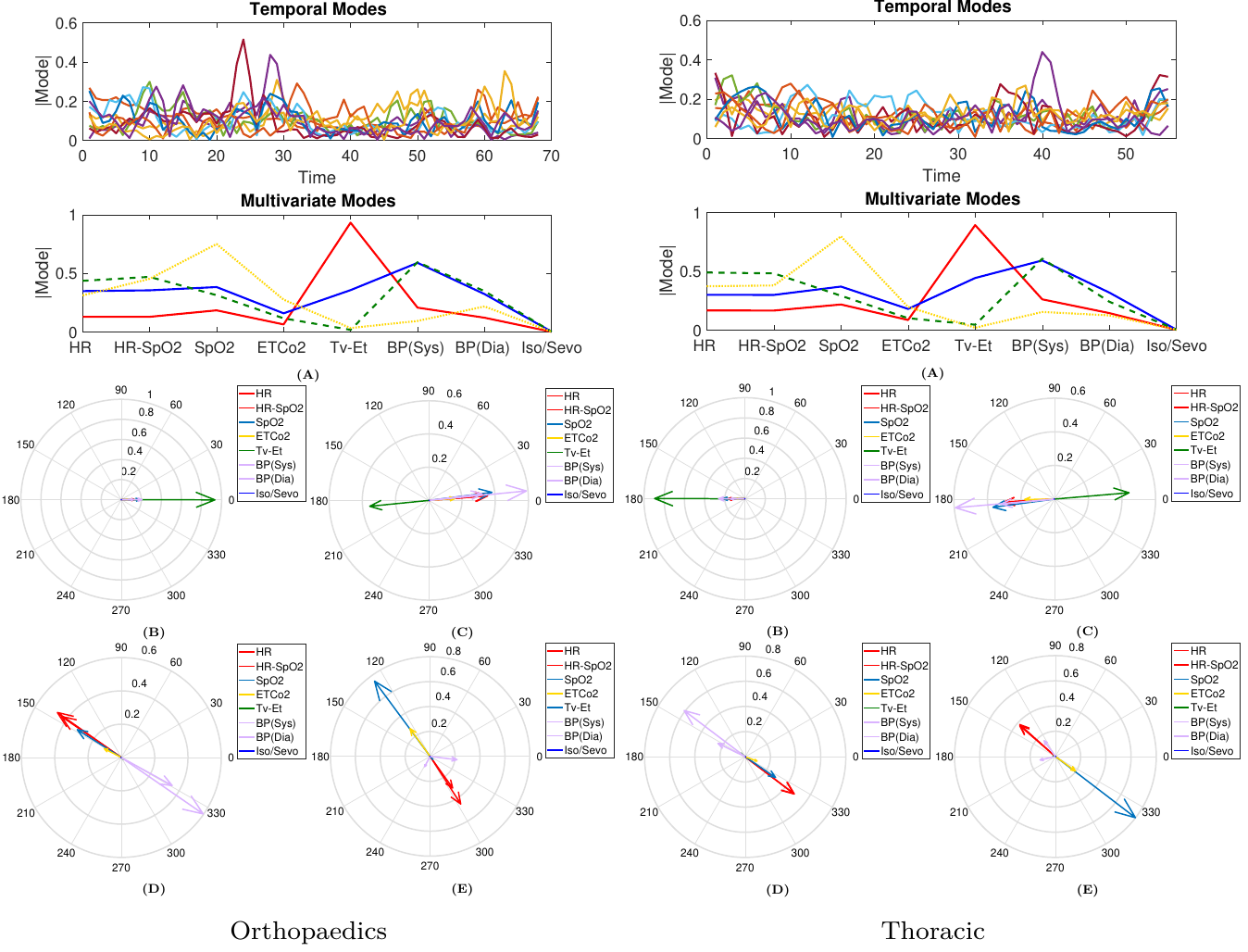}}
\caption{Comparison of the multivariate and temporal factors extracted through the complex-HOSVD for two different surgical services. A)The outer product of the temporal and multivariate factors generates the contributing components in the decomposition of $\mathcal{X}$. For intra-operative vital signs, each of the elements of a multivariate mode represents a particular vital sign. The magnitude and phase of the element explain how the vital signs are related to each other within that factor. The phase of each element describes the relative phase of the vital sign's vibration relative to the other vital signs for that multivariate factor.  This depiction allows for interpretation of the complex-HOSVD output for intra-operative vital signs: each multivariate factor identifies the vital signs involved in that pattern of physiological response in addition to the relative phase of that vital sign's activation time. B, C, D, and E) in both Orthopaedics and Thoracic surgeries represent the phase portrait associated with the multivariate factors shown in red, blue, green, and yellow, respectively. } 
\label{Fig:8}
\end{figure*}

The complex-HOSVD characterized surgical dynamics over a wide range of surgical services. Remarkably, the complex-HOSVD extracted only a few number (four) of multivariate factors to capture both the within-patients and the across-patients intra-operative dynamics. Temporal evolution of these multivariate factors are captured in 32 temporal factors and shows substantially different characteristics potentially affected by surgical services. We visualized the multivariate and temporal factors of the complex-HOSVD in figure \ref{Fig:3}. The first multivariate factor (red in figure \ref{Fig:3} A) indicates strong contribution of tidal volume for this factor. As stated in previous section, this variable is selected by anesthesiologist during surgery and impacts on hemodynamic responses. We observed that the corresponding elements of this mode oscillating with the same phase (Figure \ref{Fig:3} B).  The second multivariate mode (Blue in figure \ref{Fig:3} A) emphasizes on the contribution of heart rate and blood pressure for this factor. According to its phase plot (Figure \ref{Fig:3} C) all of the elements of this factor have the same phase except the element associated with tidal volume's participation. The third multivariate mode (Green in figure \ref{Fig:3}) indicates slightly different elemental's participation to that of indicated by the second factor. The phase information for this multivariate mode indicates a phase difference between blood pressure and the other elements of this mode (Figure \ref{Fig:3} D). Finally, the fourth multivariate mode (Yellow in figure \ref{Fig:3} A) highlights the contribution of peripheral capillary oxygen saturation (SpO2), and end-tidal Co2 while also  reveals the phase difference between these two with the other contributing elements for this mode (Figure \ref{Fig:3} E).   
%
\subsection{Grouping Patients Based on Surgical Service, Discovery of Surgery Related Features}
Every surgical procedure consists of a physical intervention on a particular body system. Hence, the type of procedure specifies the organ, organ system, or tissue involved; as well as the degree of invasiveness. The influence of the type of surgery on the development of chronic post-operative pain is well established. Longer and more complicated operations are often linked with higher risks of chronic pain development, although the pattern is irregular and also tied to the type of tissue involved in the surgery. In our analysis, the evolutionary dynamics of intra-operative vital signs have a temporal factor significantly affected by the type of surgery. Therefore, in this section, we abstract the patients into subgroups related to different surgical procedures and investigate surgery-related features (SRF) associated with the development of long-term post-operative pain. Surgery-related features may correspond to a power increase or decrease distributed over multiple intra-operative vital signs as well as changes in activation of multivariate factors' oscillating frequencies expressed by temporal modes.

This time, the input to the Complex-HOSVD algorithm is built from time-varying contents of seven intra-operative vital signs with a length of 50 minutes (started 10 minutes before incision time during surgery). By decreasing the length of intra-operative vital signs, we increased the number of patients in each subgroup. We abstracted 242 patients into six groups based on the surgical service they received. The surgical groups include thoracic, orthopedics, urological, colorectal, transplant, and pancreas \& biliary surgeries. 
We attempted to identify how surgical mechanisms in different surgeries emerge over the process of physiological responses. Here, we show the Complex-HOSVD can characterize surgery-related dynamics using physiological responses of a group of patients who received the same surgical service. Figure \ref{Fig:8} indicates the surgical dynamics characterized through the complex-HOSVD for two types of surgery (Orthopaedic, and thoracic). One more time, the complex-HOSVD summarized both the within-patients physiological responses and the across-patients dynamics in a few (four) multivariate factors. The factors offer slightly different elementals' contributions while they show the same relative phase portrait.  Apparently, the time course of these factors is substantially different over different types of surgeries. In essence, the multivariate factors indirectly encode the sympathetic activities to compensate for variations in hemodynamic parameters (autonomic regulation) under general anesthesia, and these signatures are modulated by the physiological state during surgery (captured by temporal factors).

\section{Physiological responses during surgery and Post-operative pain}
The complex principal multivariate and temporal factors extracted through the complex-HOSVD were used as new bases to describe the physiological dynamic correlations and to gain insight into any lead-lag relations among individual responses expressed in instantaneous phases of the complex vital signs. We abstracted 242 patients into two groups based on verbal evaluation of average pain on Day-30 and Day-90 after surgery. Patients reporting average pain intensity of 3 or less were categorized as “mild.” Patients reporting average pain intensity greater than 3 were considered as “severe”. The subspace provided by the complex-HOSVD can be used directly in a classification task. However, the corresponding bases do not contain any category information that is functional to model the dissimilarity among categories of data. To obtain the most salient multivariate, and temporal factors for this classification task, we used a rank feature method based on Fisher ranking, and the top 3 components were selected. The projection is performed to a 3-dimensional data manifold, which in our study are the top 3 dimensions providing the highest Fisher scores. The phase information of the projected data points is used to classify “mild” versus “severe” classes on post-operative Day-30 and Day-90 through linear discriminant analysis (LDA).
\subsection{Results for Post-Operative Day-30}
We investigated performance of the LDA in a 5-fold cross-validation procedure. The method achieved the true positive rate and positive predictive value of 0.69 and 0.60 for thoracic surgery, 0.77 and 0.67 for orthopedic surgery, 1 and 0.75 for transplant surgery, and 0.63 and 0.71 for urological surgery, respectively. In contrast, the positive predictive value and the true positive rate for the class “severe” in pancreas and colorectal surgeries reported as 0.44 and 0.57 in pancreas surgery, 0.43 and 0.86 in colorectal surgery, respectively. The results are summarized in table \ref{tab:Time:2}.
\begin{figure}[!h]
\centering
\includegraphics[width=0.49\textwidth]{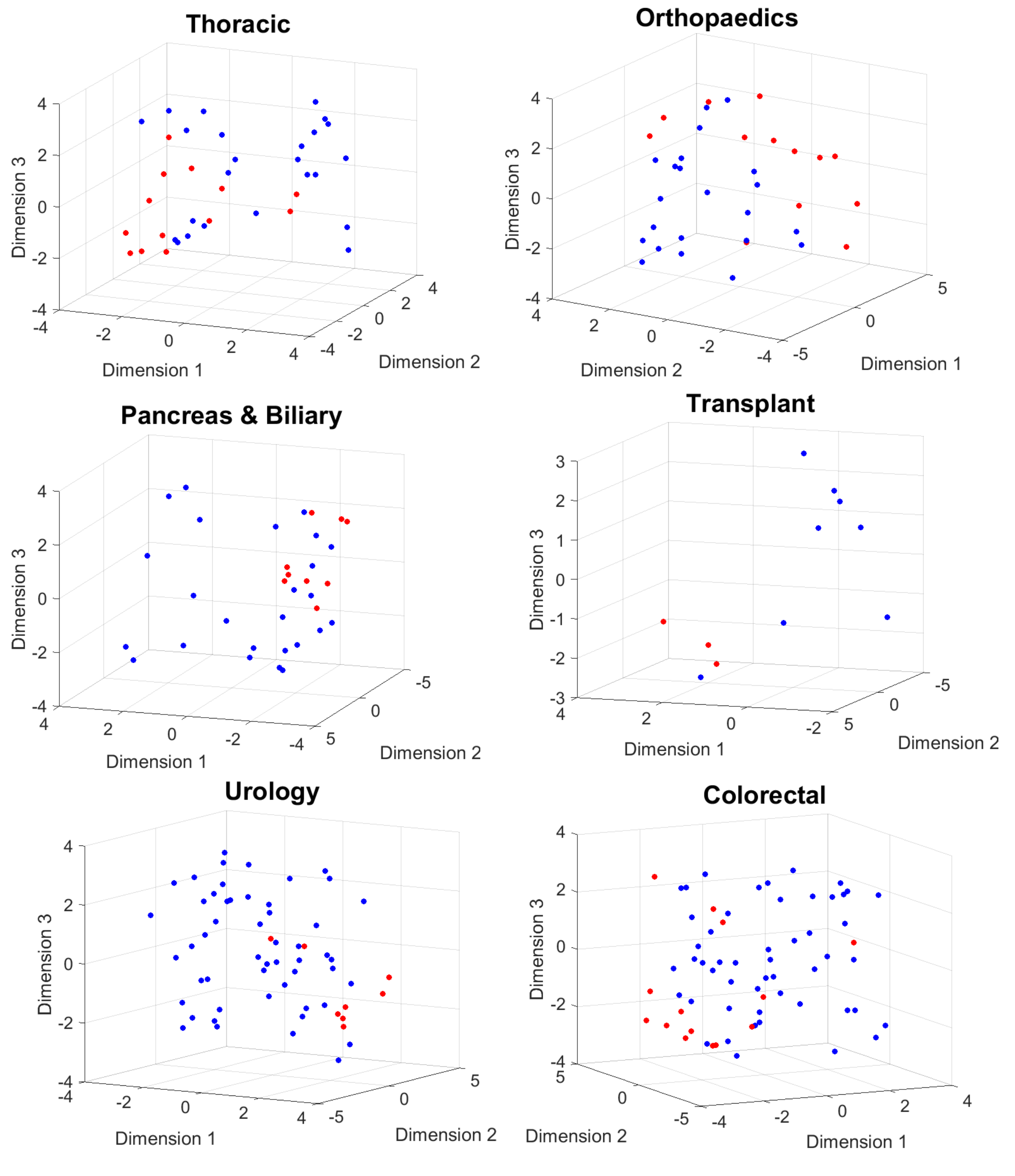}

\caption{The phase information of the projected data points onto a 3-dimensional manifold extracted using the complex-HOSVD. Mild (blue dots) versus severe (red dots) levels of pain on Day 30 after surgery is considered for this plot.} 
\label{Fig:ScatterMonth1_SurgicalServicesInTime_NotRotated}
\end{figure}
Figures \ref{Fig:ScatterMonth1_SurgicalServicesInTime_NotRotated} and \ref{Fig:ScatterMonth3_SurgicalServicesInTime_NotRotated} show the scatter plot of the phase information for patients projected onto the 3-dimensional subspace for different surgical groups. Patients with severe pain on post-operative Day-30 are almost well clustered for thoracic, orthopedics, transplant, and colorectal surgical groups. This indicates that the dynamics of patients’ physiological responses to surgical stimulation are linked to long-term post-operative pain development. Many of the same pain-category patients respond to surgical stimulation with a small band of variation of their phases. The phenomena is even better captured for severe levels of pain on post-surgical Day-90 (Figure \ref{Fig:ScatterMonth3_SurgicalServicesInTime_NotRotated}).  
\begin{table}[h!]
\caption{Performance of LDA to discriminate severe versus mild pain categories. The phase information of the projected data points onto a 3-dimensional manifold is used in the experiments. Patients are categorized based on their surgical service. Month 1 without rotation}
\setlength{\tabcolsep}{3pt}
      \begin{tabular}{p{46pt} p{59pt} p{31pt} p{36pt} p{36pt} p{15pt} }
      \hline
           Surgery & Confusion matrix  &Precision  & Sensitivity & Specificity& AUC\\ 
           &   &(PPV)   & (TPR) & (TNR)& \\ \hline
        & TP=9\quad FN=4 &\textbf{0.60} & \textbf{0.69}& 0.75&0.78 \\
         \textbf{Thoracic}  &\enspace FP=6\quad TN=18&  & & & \\ \hline
          
         & TP=10\quad FN=3&\textbf{0.67}   & \textbf{0.77} & 0.77&0.80 \\
         \textbf{Orthopaedics} &\enspace FP=5\quad TN=17 &   &  & & \\ \hline
                   
                    & TP=5\quad FN=3& \textbf{0.71} & \textbf{0.63} & 0.96&0.87 \\
          \textbf{Urology} &\enspace FP=2\quad TN=50 &   &  & & \\ \hline
                    
                   & TP=6\quad FN=8 & \textbf{0.86}& 0.43&0.98&0.75 \\
         \textbf{Colorectal} &\enspace FP=1\quad TN=50&  & & & \\ \hline
          
         & TP=3\quad FN=0&\textbf{0.75}   &\textbf{ 1} &0.88 &0.92 \\
         \textbf{Transplant} &\enspace FP=1\quad TN=7 &   &  & & \\ \hline
                   
                   & TP=4\quad FN=5&0.57  & 0.44 & 0.88&0.80 \\
          \textbf{Pancreas \& Biliary} &\enspace FP=3\quad TN=22 &   &  & & \\              
                    \hline      
      \end{tabular}
      \label{tab:Time:2}
\end{table}
\subsection{Results for Post-Operative Day-90}
Given that healing times vary between different procedures, and the duration for persistent post-operative pain is defined at 3 months after surgery by the  International Classification of Diseases, we repeated the exact set of experiments to classify the patients who reported "mild" versus "severe" levels of pain on post-operative day-90 as well. We observed that although the number of patients included in the "severe" class decreased for all surgical groups we achieved almost the same or higher performances in detecting the patients who developed severe versus mild levels of pain  (except for urological and orthopedics surgeries). The results are summarized in table \ref{tab:Time:3}.
\begin{figure}[h!]
\centering
\includegraphics[width=0.49\textwidth]{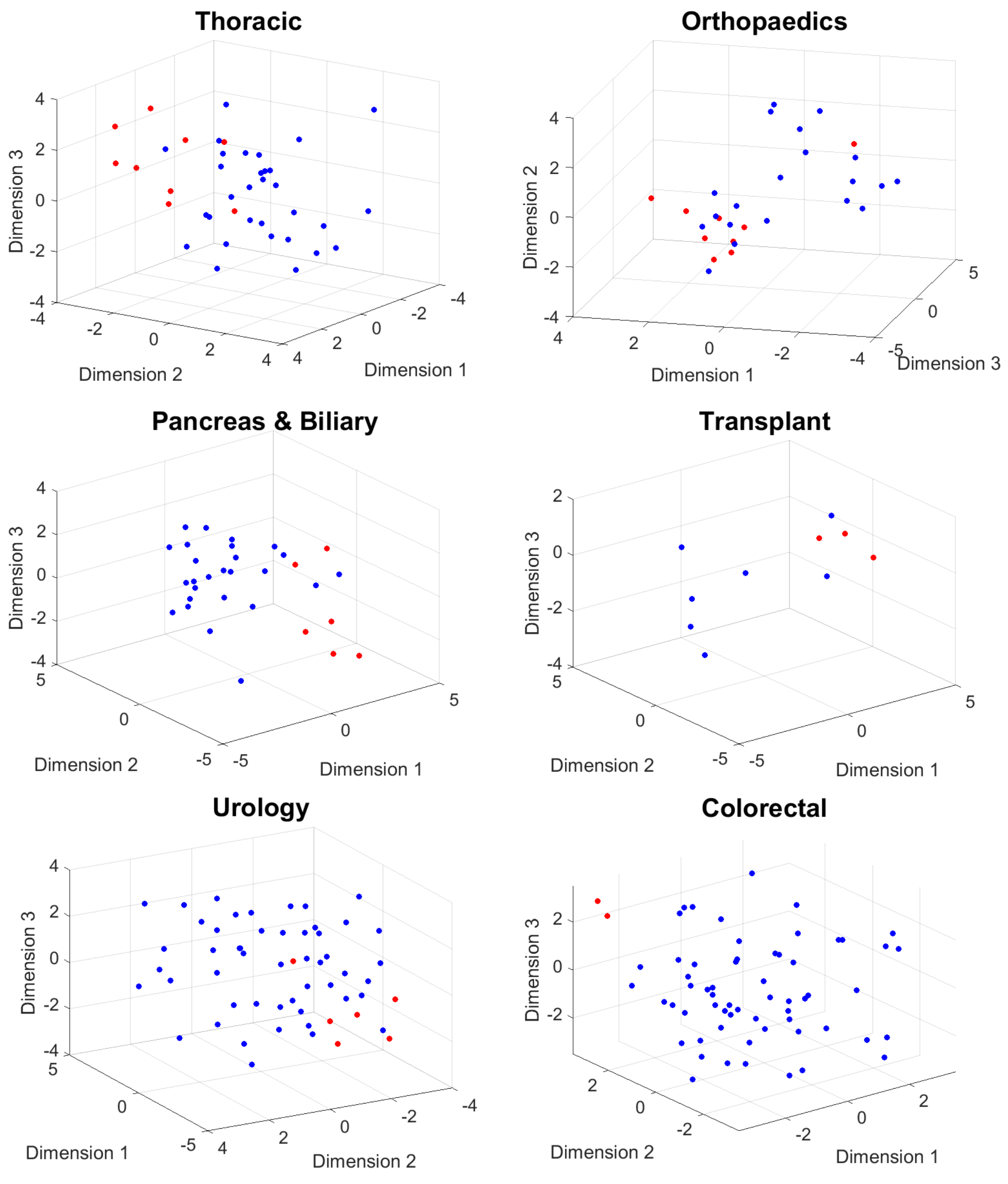}
\caption{The phase information of the projected data points onto a 3-dimensional manifold extracted using the complex-HOSVD. Mild (blue dots) versus severe (red dots) levels of pain on Day 90 after surgery is considered for this plot. } 
\label{Fig:ScatterMonth3_SurgicalServicesInTime_NotRotated}
\end{figure} 
\begin{table}[h!]
\caption{Performance of LDA to discriminate severe versus mild pain categories. The phase information of the projected data points onto a 3-dimensional manifold is used in the experiments. Patients are categorized based on their surgical service. Month 3 without rotation }
\setlength{\tabcolsep}{3pt}
      \begin{tabular}{p{46pt} p{59pt} p{31pt} p{36pt} p{36pt} p{15pt}}
        \hline
           Surgery & Confusion matrix  &Precision  & Sensitivity & Specificity& AUC\\ 
           &   &(PPV)   & (TPR) & (TNR)& \\ \hline
        & TP=6\quad FN=3 &\textbf{0.75} & \textbf{0.67}& 0.94&0.87 \\
         \textbf{Thoracic}  &\enspace FP=2\quad TN=29&  & & & \\ \hline
          
         & TP=6\quad FN=3&\textbf{0.55}   & \textbf{0.67} & 0.75&0.73 \\
         \textbf{Orthopaedics} &\enspace FP=5\quad TN=15 &   &  & & \\ \hline
                   
                    & TP=2\quad FN=4& \textbf{0.67} & \textbf{0.33} & 0.98&0.88 \\
          \textbf{Urology} &\enspace FP=1\quad TN=49 &   &  & & \\  \hline
                    
                   & TP=2\quad FN=0 & \textbf{1}& 1&1&1 \\
         \textbf{Colorectal} &\enspace FP=0\quad TN=60&  & & & \\ \hline
          
         & TP=2\quad FN=1&\textbf{0.67}   &\textbf{ 0.67} &0.86 &0.90 \\
         \textbf{Transplant} &\enspace FP=1\quad TN=6 &   &  & & \\ \hline
                   
                   & TP=4\quad FN=2&0.67 &0.67 & 0.92&0.92 \\
          \textbf{Pancreas \& Biliary} &\enspace FP=2\quad TN=24 &   &  & & \\              
                    \hline      
      \end{tabular}
      \label{tab:Time:3}
\end{table}
\\
\indent 
\begin{figure*}[h!]
\includegraphics[width=0.9\textwidth]{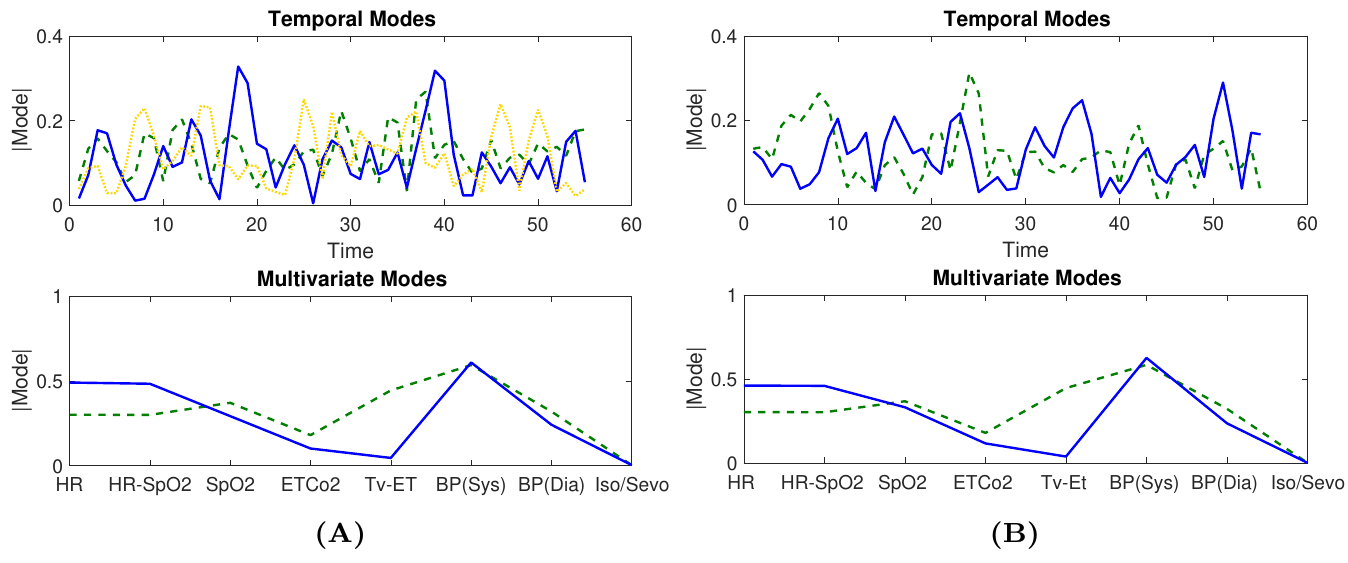}
\centering
\caption{Thoracic surgery: The contributing multivariate-temporal factors (blue, green, and yellow indicate rank from the highest to the lowest). A)Post-surgical Day-30:The multivariate modes show that the most dissimilarities between "mild" versus "severe" levels of pain are encoded in variations of heart rate and blood pressure (Blue and Green vectors). Time evolution of the multivariate modes are encoded in temporal modes. B)Post-surgical Day-90: The two multivariate factors (Blue and green vectors) emphasize on strong contribution of heart rate and blood pressure while they have distinct phase portraits (Figure \ref{Fig:8}). } 
\label{Fig:12}
\end{figure*}
Figure \ref{Fig:12} compares the contributing multivariate-temporal factors for the first three leading components with the highest Fisher scores in differentiation between "mild" versus "severe" levels of pain on post-surgical Day-30 and Day-90 for thoracic surgery. For post-surgical Day-30:  The first and the second multivariate factors emphasize the role of heart rate and blood pressure. The activation of the second multivariate mode (green vector) was captured in two distinct temporal factors (green and yellow vectors). Figure \ref{Fig:ScatterMonth1_SurgicalServicesInTime_NotRotated} shows that almost all patients who developed severe levels of pain on post-operative Day-30 have negative phases with respect to first and second dimensions, mostly focused on changes in heart rate and blood pressure. For post-surgical Day-90:  The first and the second multivariate factors emphasize the role of heart rate and blood pressure. The activation of the first multivariate mode (blue vector) was captured in two distinct temporal factors (blue and green vectors). Figure \ref{Fig:ScatterMonth3_SurgicalServicesInTime_NotRotated} shows that almost all patients who developed severe levels of pain on post-operative Day-90 have positive phases with respect to the first and the third dimensions, and negative phases with respect to the second dimension.
\section{Rotating the physiological responses with respect to patients' dynamic variation}
As discussed earlier, each complex-HOSVD component identifies sub-physiological parameters (multivariate factor), with common intra-surgery temporal dynamics (temporal factor), which were deferentially activated across patients. Overall, the complex-HOSVD model uncovered a reasonable portrait of surgical dynamics (population dynamics) in which distinct subsets of physiological parameters are active at different times during surgery and whose variation across patients encoded in individual dynamic variables.\\
\indent
\begin{figure*}[!h]
\centering
\includegraphics[width=0.7\textwidth]{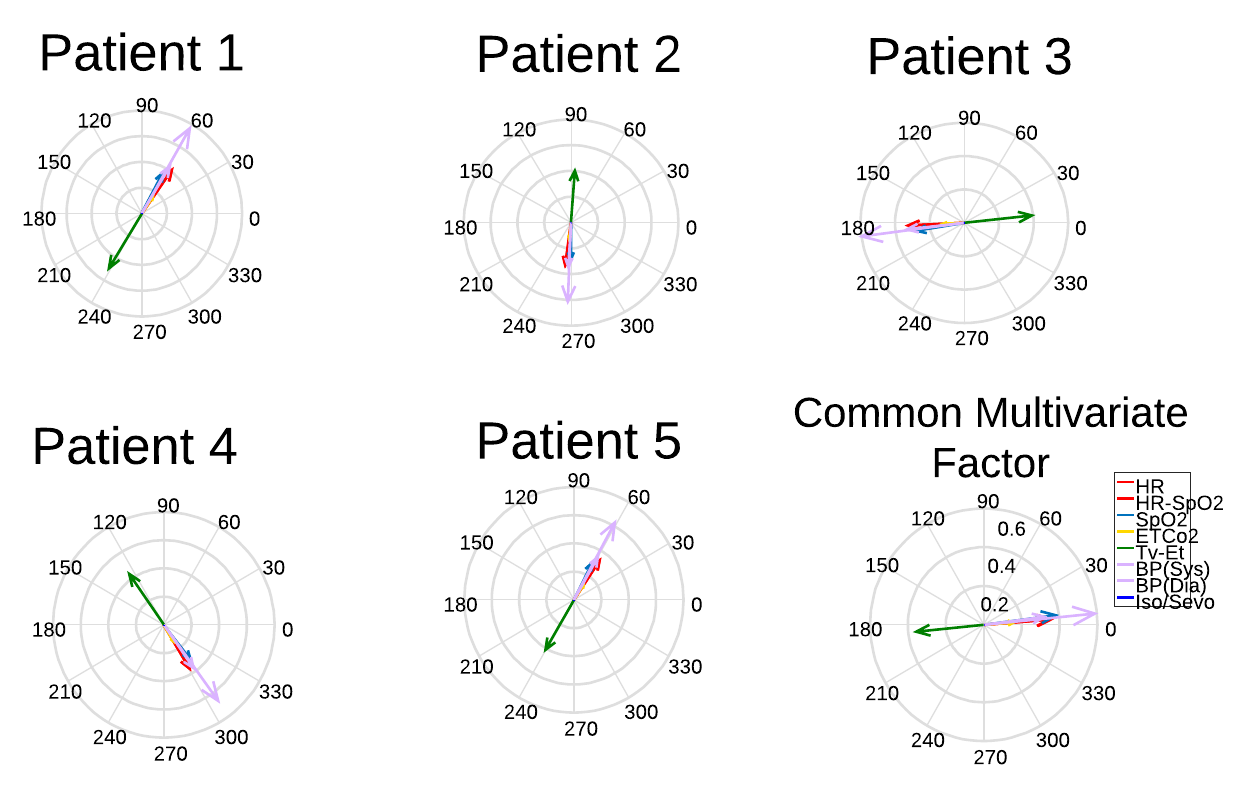}
\caption{The multivariate dynamic variation for different patients in Transplant surgery. For intra-operative vital signs, each of the elements of a multivariate mode represents a particular vital sign. The magnitude and phase of the element explain how the vital signs are related to each other within that factor. The phase of each element describes the relative phase of the vital sign's vibration relative to the other vital signs for that multivariate factor.  Although the elements of the multivariate factors for different patients have the same relative phase the dynamics are not exactly aligned.} 
\label{Fig:dynamic changes}
\end{figure*}
Until now, we used the common multivariate-temporal dynamics as new bases to describe physiological responses, hence we discarded individual dynamic variations encoded in "patients factors". However, for a better representation of dynamics, it is essential to associate each principal component (as one base of the subspace) to each dynamic mode of patients' physiological responses ~\cite{andalib2018framework,andalib2019unsupervised}. The coordinate systems provided by the common multivariate-temporal factors and the patients' multivariate-temporal dynamics are not necessarily the same (are not aligned exactly) ~\cite{andalib2018framework,andalib2019unsupervised}. Given that all factors extracted through the complex-HOSVD are complex-valued factors, the patients-specific variations for the multivariate-temporal dynamics identified by the method contain scaling and rotational adjustments that appeared in the outer product of the multivariate-temporal dynamics with the patients' factors. Figure \ref{Fig:dynamic changes} shows how multivariate dynamic changes across 5 patients in Transplant surgery. For simplicity temporal factors are discarded in this figure, but the same adjustments apply for temporal factors as well.
\indent
\begin{figure*}[!h]
\centering
\includegraphics[width=0.7\textwidth]{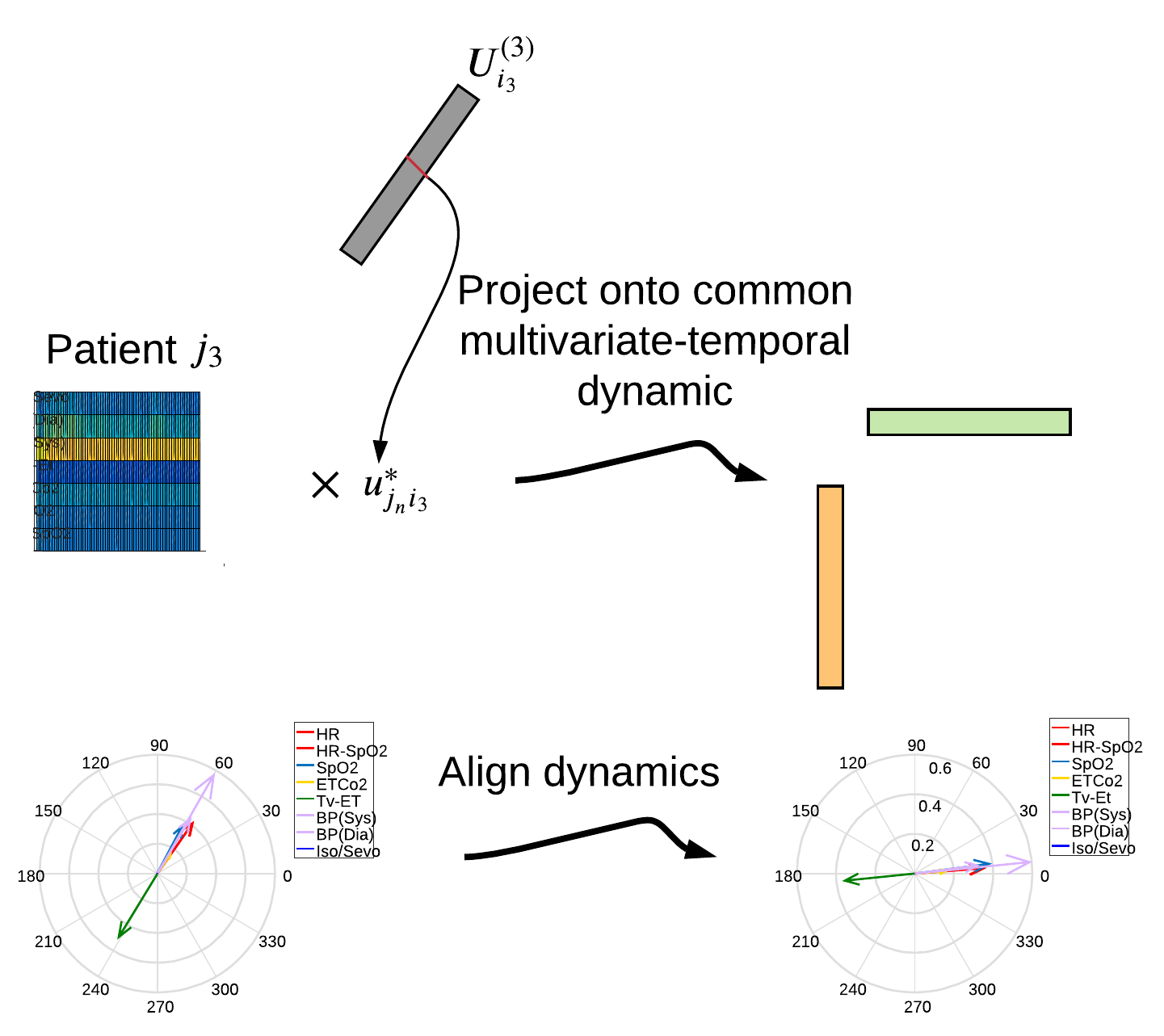}
\caption{Rotation of physiological responses before projection to align with the common multivariate-temporal dynamic. for simplicity, one component is shown here.} 
\label{Fig:Rotation by complex conjugate}
\end{figure*}

To compare the complex correlations between each physiological response and the extracted multivariate-temporal dynamics, it is essential to have a common coordinate system for all patients. Simultaneously, to account for dynamic variation across patients, instead of rotating the dynamics, the complex conjugate of elements, given by the patients' factors, may be used to scale and rotate the physiological responses before projection onto the subspace. The process can be done per complex-HOSVD component separately. From a geometrical point of view, the process can be considered as an active transformation in which the position of a point changes in a coordinate system; whereas a passive transformation changes the coordinate system in which the point is described. Figure \ref{Fig:Rotation by complex conjugate} shows how the process works. 

Once the new projections were obtained we repeated the same set of experiments to explore dynamic correlations in intra-operative vital signs. Again the phase information of the projected data points is used to classify "mild" versus "severe" classes on post-operative Day-30 and Day-90 through linear discriminant analysis. 

\subsection{Results for Post-Operative Day-30}
\begin{figure}[h!]
\includegraphics[width=0.49\textwidth]{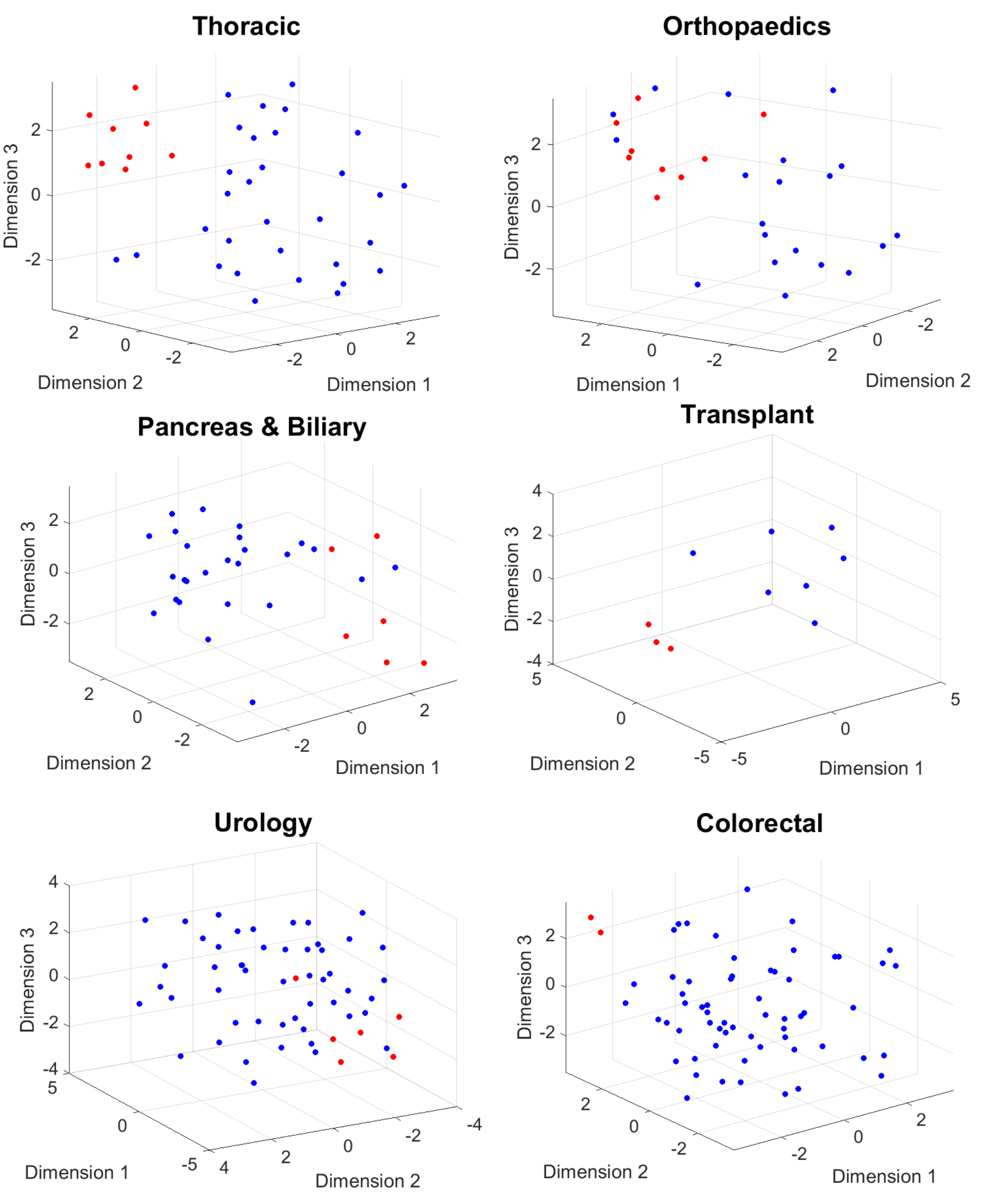}
\caption{The phase information of the projected data points onto a 3-dimensional manifold extracted using the complex-HOSVD. Mild (blue dots) versus severe (red dots) levels of pain on Day 90 after surgery is considered for this plot. } 
\label{Fig:ScatterMonth3Rotated}
\end{figure}

We noticed that the true positive rate, or the positive predictive value for the class of severe pain in five groups related to Thoracic, Orthopaedics, Colorectal, Transplant, and Pancreas \& Biliary surgeries improved compared with the results of the previous section. The true positive rate and positive predictive values were 0.69 and 0.75 for thoracic surgery, 0.77 and 0.83 for orthopedic surgery, 1 and 1 for transplant surgery,  0.57 and 0.73 for colorectal surgery, and 0.67 and 0.67 for Pancreas \& Biliary surgery, respectively. In contrast, the positive predictive value and the true positive rate for the class “severe” in urological surgery remained the same. The results are summarized in table \ref{tab:Time:4}.
\begin{figure}[!h]
\includegraphics[width=0.49\textwidth]{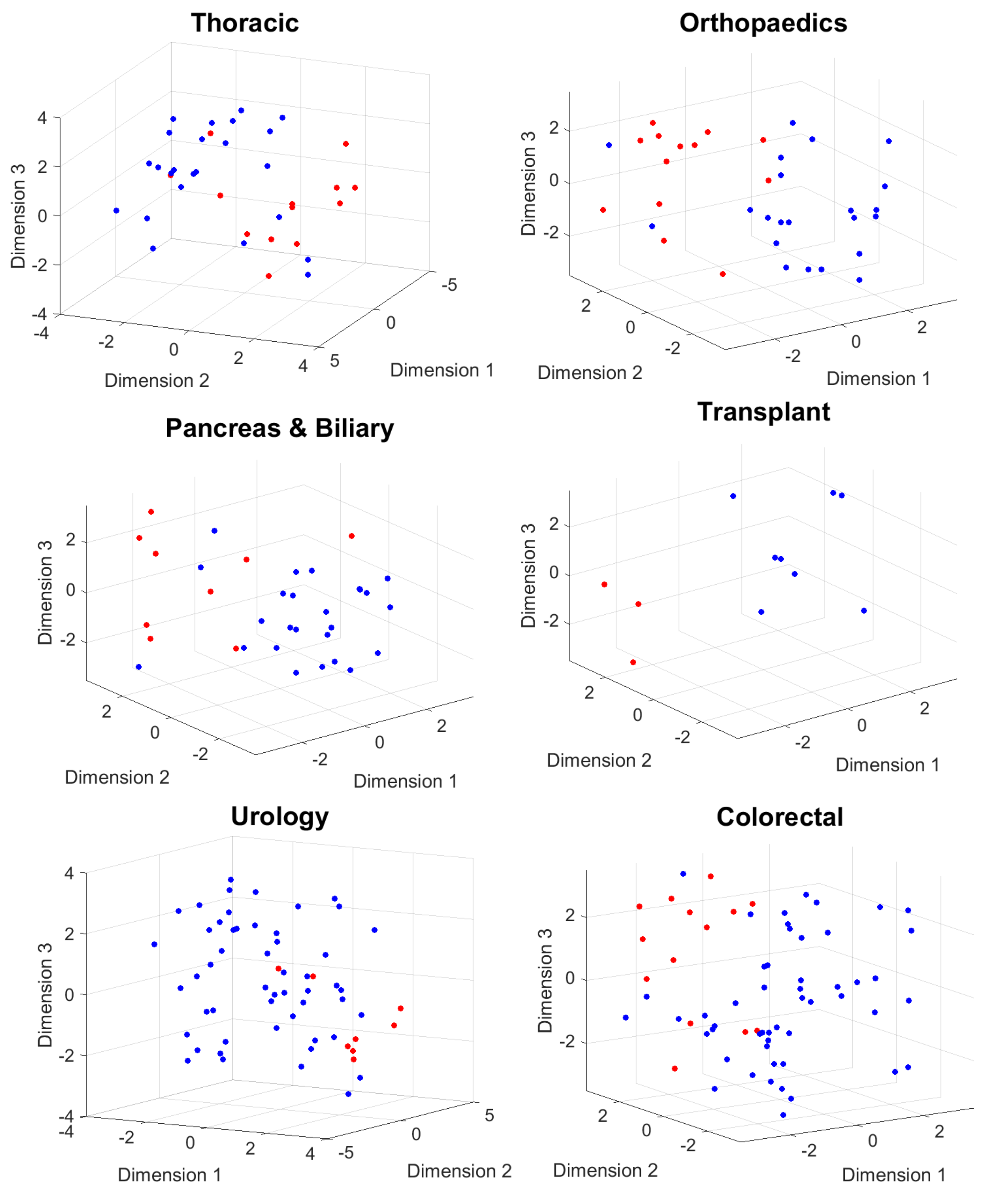}
\caption{The phase information of the projected data points onto a 3-dimensional manifold extracted using the complex-HOSVD. Mild (blue dots) versus severe (red dots) levels of pain on Day 30 after surgery is considered for this plot. } 
\label{Fig:Month1Rotated}
\end{figure}
Figures \ref{Fig:Month1Rotated} and \ref{Fig:ScatterMonth3Rotated} show the scatter plot of the phase information for patients projected onto the 3-dimensional subspace for different surgical groups. The phase information of the physiological responses of the patients in the same group of pain is more similar to each other than to those in the other group in thoracic, orthopaedics, transplant, pancreas \& biliary, and colorectal surgical groups. This emphasizes again that the dynamics of patients’ physiological responses to surgical stimulation are associated with long-term post-operative pain development. We observed the same pattern in the results for post-surgical Day-90 (Figure \ref{Fig:ScatterMonth3Rotated}). 
\begin{table}[!h]
\caption{Performance of LDA to discriminate severe versus mild pain categories. The phase information of the projected data points onto a 3-dimensional manifold is used in the experiments. Patients are categorized based on their surgical service. The physiological responses are rotated before projection.}
\setlength{\tabcolsep}{3pt}
      \begin{tabular}{p{46pt} p{59pt} p{31pt} p{36pt} p{36pt} p{15pt}}
      \hline
           Surgery & Confusion matrix  &Precision  & Sensitivity & Specificity& AUC\\ 
           &   &(PPV)   & (TPR) & (TNR)& \\ \hline
        & TP=9\quad FN=4 &\textbf{0.75} & \textbf{0.69}& 0.88&0.81 \\
         \textbf{Thoracic}  &\enspace FP=3\quad TN=21&  & & & \\ \hline
          
         & TP=10\quad FN=3&\textbf{0.83}   & \textbf{0.77} & 0.91&0.87 \\
         \textbf{Orthopaedics} &\enspace FP=2\quad TN=20 &   &  & & \\ \hline
                   
                    & TP=5\quad FN=3& \textbf{0.71} & \textbf{0.63} & 0.96&0.87 \\
          \textbf{Urology} &\enspace FP=2\quad TN=50 &   &  & & \\ \hline
                    
                   & TP=8\quad FN=6 & \textbf{0.73}& 0.57&0.94&0.86 \\
         \textbf{Colorectal} &\enspace FP=3\quad TN=48&  & & & \\ \hline
          
         & TP=3\quad FN=0&\textbf{1}   &\textbf{ 1} &\textbf{ 1} &1 \\
         \textbf{Transplant} &\enspace FP=0\quad TN=8 &   &  & & \\ \hline
                   
                   & TP=6\quad FN=3&0.67  & 0.67 & 0.88&0.83 \\
          \textbf{Pancreas \& Biliary} &\enspace FP=3\quad TN=22 &   &  & & \\              
                    \hline

      \end{tabular}
      \label{tab:Time:4}
\end{table}

\subsection{Results for Post-Operative Day-90}
Again, we repeated the exact set of experiments to classify the patients who reported "mild" versus "severe" levels of pain on post-operative Day-90. We observed that the true positive rate, and the positive predictive value for the class of severe pain in three groups related to Thoracic, Orthopaedics, and Transplant increased. The results are summarized in Table \ref{tab:Time:5}.
 
\begin{table}[h!]
\caption{Performance of the LDA to discriminate severe versus mild pain categories. The phase information of the projected data points onto a 3-dimensional manifold is used in the experiments. Patients are categorized based on their surgical service. The physiological responses are rotated before projection.}
\setlength{\tabcolsep}{3pt}
      \begin{tabular}{p{46pt} p{59pt} p{31pt} p{36pt} p{36pt} p{15pt} }
      \hline
           Surgery & Confusion matrix  &Precision  & Sensitivity & Specificity& AUC\\ 
           &   &(PPV)   & (TPR) & (TNR)& \\ \hline
        & TP=8\quad FN=1 &\textbf{0.80} & \textbf{0.89}& 0.94&0.89 \\
         \textbf{Thoracic}  &\enspace FP=2\quad TN=29&  & & & \\ \hline
          
         & TP=7\quad FN=2&\textbf{0.64}   & \textbf{0.78} & 0.80&0.83 \\
         \textbf{Orthopaedics} &\enspace FP=4\quad TN=16 &   &  & & \\ \hline
                   
                    & TP=2\quad FN=4& \textbf{0.67} & \textbf{0.33} & 0.98&0.88 \\
          \textbf{Urology} &\enspace FP=1\quad TN=49 &   &  & & \\ \hline
                    
                   & TP=2\quad FN=0 & \textbf{1}& 1&1&1 \\
         \textbf{Colorectal} &\enspace FP=0\quad TN=60&  & & & \\ \hline
          
         & TP=3\quad FN=0&\textbf{1}   &\textbf{ 1} &1 &1 \\
         \textbf{Transplant} &\enspace FP=0\quad TN=7 &   &  & & \\ \hline
                   
                   & TP=4\quad FN=2&0.67  & 0.67 & 0.92&0.92 \\
          \textbf{Pancreas \& Biliary} &\enspace FP=2\quad TN=24 &   &  & & \\              
                    \hline        
      \end{tabular}
      \label{tab:Time:5}
\end{table}

\section{Discussion}
This study introduced a new type of multivariate-temporal decomposition of intra-operative vital signs to explore signatures that discriminate patients who develop mild or severe pain on post-operative Day-30 and Day-90. The method takes advantage of the fact that the complex-HOSVD decomposes data into a sum of rank-1 tensors which is a combination of modes or signatures. The method arranges the multivariate trajectory of intra-operative vital signs of various patients in a three-dimensional data array with dimensions indexed by vital sign variable, time, and patient. This is the first time that multivariate-temporal decomposition of complex-valued intra-operative vital signs is proposed to analyze long-term post-operative pain. Employing a multivariate-time structure helped us to accurately describe the dynamics of intra-operative vital signs and to find a lower-dimensional projection where differences between individual responses were encoded in the phases of the complex vital signs.

The primary advantage of the complex-HOSVD is that it discovers and examines multivariate-temporal behavior. However, the complex multivariate-temporal factors are difficult to interpret as amplitude and phase relations due to Hilbert transform properties, which weigh more sudden transitions than the episodes during which the intra-operative vital signs change slowly.  Further work is necessary to compensate for this behavior. 

This study presents a physiological interpretation of the model, although clinical verification has not yet been undertaken. The interpretation focused on the spectral dynamics of different vital signs during surgery. For the intra-operative vital signs time series used in this study, the spectral band is within the frequency band of the autonomic nervous system responding to surgical stimulus under general anesthesia which established the sampling rate. 

Variability in physiological parameters during surgery is a result of dynamic interaction between surgical induced perturbations to the circulatory system and the short-term compensatory response to regulate them. Short-term circulation control by the baroreceptor reflex or vasomotor tone, for instance, is best described by feedback models. The circulation control could be identified by the pair of input-output signals. In the case of baroreceptor reflex, blood pressure and heart rate act as input and output signals, respectively. Transfer function parameters in the feedback system determine the input-output relation.  While the gain defines the input-output signals' amplitude relationship, the phase determines the delay between the two. In the baroreceptor reflex case, the phase of transfer function quantifies the phase shift between blood pressure and heart rate. We can think of multivariate factors as prototypical short-term circulation control patient patterns. Hence, the complex-valued elements of the multivariate factors might correspond to the attributes of the transfer function. In this setting, the absolute value of elements might correlate with the gain of the transfer function and the angle might indicate the delay between input and output signals. For example, the strong contribution of the heart rate and blood pressure and the phase shift between them in one of the extracted multivariate factors might correspond to circulation control by baroreceptor reflex (Figure \ref{Fig:3} D).

Temporal factors are highly surgical-dependent, hence more difficult to interpret. However, they may serve as indicators of circulation control activity during surgery. 

Our study is limited by the sampling rate of intra-operative vital signs, verbal evaluation of post-operative pain, and the limited number of surgical patients involved. The higher sampling rate of vital signs would allow a more comprehensive analysis of autonomic nervous system activity. A larger number of patients would provide valid testing of various hypotheses about temporal and multivariate factors within surgeries. Finally, a more reliable method to assess post-operative pain would remove noise from our data set.

\bibliography{myUnifiedRef}
\bibliographystyle{IEEEtran}

\end{document}